\documentclass[pra,aps,twocolumn,amsmath,amssymb]{revtex4}

\usepackage{graphicx}
\usepackage{dcolumn}
\usepackage{bm}
\usepackage[extension=xxx]{hyperref}

\newcommand{\beq}{\begin{equation}}
\newcommand{\eeq}{\end{equation}}
\newcommand{\beqa}{\begin{eqnarray}}
\newcommand{\eeqa}{\end{eqnarray}}

\def\beq{\begin{equation}}

\usepackage{color}

\begin{document}

\title{Resonant excitations of a Bose Einstein condensate in an optical lattice}
\date{\today}

\author{C. Cabrera-Guti\'errez$^1$, E. Michon$^1$, M. Arnal$^1$, V. Brunaud$^1$, T. Kawalec$^2$, J. Billy$^1$, D. Gu\'ery-Odelin$^1$}

\affiliation{$^1$ Universit\'e de Toulouse ; UPS ; Laboratoire Collisions Agr\'egats R\'eactivit\'e, IRSAMC ; F-31062 Toulouse, France} 
\affiliation{CNRS ; UMR 5589 ; F-31062 Toulouse, France}
\affiliation{$^2$ Marian Smoluchowski Institute of Physics, Jagiellonian University, \L{}ojasiewicza 11, PL-30348 Krak\'ow, Poland} 
\begin{abstract}
We investigate experimentally a Bose Einstein condensate placed in a 1D optical lattice whose phase or amplitude is modulated in a frequency range resonant with the first bands of the band structure. We study the combined effect of the strength of interactions and external confinement on the 1 and 2-phonon transitions. We identify lines immune or sensitive to atom-atom interactions. Experimental results are  in good agreement with numerical simulations. Using the band mapping technique, we get a direct access to the populations that have undergone $n$-phonon transitions for each modulation frequency.  \end{abstract}
\maketitle

Many recent cold atom experiments in optical lattices have demonstrated the usefulness of time-dependent modulation to probe or engineer single and many-body states \cite{ChuPRL,ArimondoPRL,Oberthaler,NJP18,RMPEckardt,am1,am2,Kollath,probe2,mode1,transport1,transport2,PRXJean,Lellouch18,Hofstadter,Haldane,sengstock}. Shaken lattices have also many applications for atom interferometry \cite{interf1,interf2,interf3}. Using modulation frequencies in a range of values different from interband resonances has been used to renormalize the tunneling rate \cite{ChuPRL,ArimondoPRL,Oberthaler,NJP18,RMPEckardt}, to probe the Mott insulator - superfluid transition \cite{am1,am2,Kollath,probe2}, to excite collective modes \cite{mode1}, to drive quantum transport \cite{transport1,transport2}, to endow a system with new properties including artificial gauge fields \cite{PRXJean,Lellouch18}, or to create nontrivial topological band structures \cite{Hofstadter,Haldane,sengstock,RMPEckardt} to name a few. 

For a large amplitude of modulation and a sufficiently deep optical lattice, the classical phase space becomes mixed. This regime has been investigated in the early 2000 to set up experiments dedicated to the observation of dynamical tunneling \cite{cat1,cat2,cat3}. For a modulation with a rich spectrum, the classical phase space even becomes fully chaotic, a regime that was used to investigate 3D Anderson localization \cite{anderson}.
 
Resonant modulation favors interband transitions \cite{raizen94,simonet,PRA2013,Interfero,Lacki} and band hybridization \cite{hy1,hy2,hy3}. It has been used as a spectroscopic calibration and for manipulation of wave packets \cite{raizen94,holder,simonet,PRA2013,arlt,spectroscopy}. Polychromatic resonant modulation has also been investigated revealing the possible interference between separated excitation paths \cite{coherentcontrol,polycexc}. However, such resonant modulation may cause heating \cite{RMPEckardt}. A detailed understanding of resonant excitations appears essential to exploit the full potentialities of Floquet engineering \cite{RMPEckardt}.

Furthermore, the band structure of a quantum gas in an optical lattice can be modified by atom-atom interactions.  When the interaction energy is on the order of the lattice depth, the band structure exhibits a loop structure about the center of the Brillouin zone allowing for dynamical instabilities and possible hysteretic behavior  \cite{loop1,loop2,loop3,loop4}. 

In this article, we investigate experimentally and numerically resonant interband excitations in a Bose-condensed atomic cloud confined by a 1D modulated optical lattice in the presence of weak interactions. By weak interactions we mean that the interaction energy is significantly lower than the lattice depth. However, we find a clear signature of interactions in our spectrum studies whilst the band structure is not significantly modified by interactions. In this work, we put the emphasis on 1 and 2-phonon transitions. 

The paper is organized as follows. In Sec.~\ref{sec:overview}, we provide an overview of the relevant features of our experimental setup. In Sec.~\ref{sec:selesctionrules}, we discuss the selection rules at work depending on the types of modulation (phase or amplitude modulation), along with their limitations. Section \ref{sec:numerics} provides the results of our numerical simulations and enables us to give some insight about the role of atom-atom interactions on the spectrum. The experimental spectra are then presented in Section \ref{sec:experiment}. In Sec.~\ref{sec:dynamics}, we detail our experimental approach and analysis of the dynamics of the excited atoms.  

\section{Overview of the lattice setup}
\label{sec:overview}

The lattice results from the interference of two counter propagating laser beams originating from the same single mode laser (wavelength 1064 nm). The excitation is carried out either by phase or amplitude modulation. The lattice potential experienced by the rubidium-87 atoms reads
\begin{equation}
V_{\varepsilon, \theta}(x,t)=-V_0 (1+\varepsilon (t)) \cos^2\left( \frac{\pi x}{d} + \theta(t) \right),
\end{equation}
where $d=532$ nm is the lattice spacing. The depth is measured in units of the characteristic energy of the lattice through the dimensionless parameter $s_0$: $V_0=s_0E_L$ with $E_L=h^2/(2md^2)$ \cite{footnote1}. The atoms also experience an external confinement $m\omega_{\rm E1,2}^2x^2/2$ provided by the hybrid trap in which the BEC is produced \cite{PRL2016,PRA2018}. Using one or two dipole beams for the hybrid trap, we have the possibility to change the angular frequency of this extra confinement by a factor 10 from $\omega_{\rm E1}=2\pi \times 10$ Hz to $\omega_{\rm E2}=2\pi \times 100$ Hz.

To engineer the phase and amplitude of the lattice, we proceed in the following manner: a 15 Watts single mode laser used for the optical lattice is first diffracted by an Acousto-Optic Modulator (AOM$_0$), and then split into two beams having the same power. Each beam is subsequently diffracted by an AOM before entering into the cell chamber. Those two latter AOMs are driven by two phase-locked synthesizers that imprint their relative phase on the laser light. The modulation amplitude, $\varepsilon(t)$, is realized by tuning the RF power of the AOM$_0$ while the phase control on $\theta(t)$ is achieved by an appropriate pre-programming of the synthesizers. 

\section{Selection rules for phase and amplitude modulation}
\label{sec:selesctionrules}

For a BEC adiabatically loaded into the static optical lattice $V_{0,0}(x) $, the wave function lies in the ground state band. Once the modulation is switched on, interband transitions can occur. The interband transition probability is proportional to the square modulus of the matrix element  
\begin{equation}
\delta V_{nn'kk'} (t)=\langle \psi_{n',k'} | \left( V_{\varepsilon, \theta}(x,t) - V_{0,0}(x)  \right) | \psi_{n,k} \rangle, 
\label{me}
\end{equation}
where $\{  |\psi_{n,k} \rangle  \}$ are the Bloch states, $n$ is the band index ($n=1,2,...$) and $k$ the quasi-momentum.  Bloch states are usually rewritten in terms of the periodic Bloch functions, $u_{n,k}(x)$, as $\psi_{n,k} (x) = e^{ikx}u_{n,k}(x)$ with $u_{n,k}(x+d)=u_{n,k}(x)$. The matrix element (\ref{me}) vanishes as soon as $k\neq k'$, because of the symmetries of the modulation potential and the properties of the Bloch function. 

Phase and amplitude modulations do not obey the same selection rules for the transfer from a band of index $n$ to $n'$. 

Consider a phase modulated optical lattice ($\varepsilon(t)=0$ and $\theta(t)=\theta_0 \sin(\omega t)$), the lattice potential reads
\begin{eqnarray}
& & V_{0, \theta}(x,t)  =   -  \frac{V_0}{2}-  \frac{V_0}{2} \bigg \{ \cos\left( \frac{2\pi x}{d}  \right)J_0(2 \theta_0) \nonumber \\
& &+  2\cos\left( \frac{2\pi x}{d}  \right) \sum_{p=1}^\infty J_{2p}(2 \theta_0) \cos(2p\omega t) \nonumber \\
& &-  2\sin\left( \frac{2\pi x}{d}  \right) \sum_{p=0}^\infty J_{2p+1}(2 \theta_0) \sin((2p+1)\omega t)   \bigg \},
 \label{eqpotmodphas}
 \end{eqnarray}
where $J_p(x)$ is the first kind Bessel function of order $p$. We first notice that the phase modulation renormalizes the depth of the lattice from $V_0$ to $V_0J_0(2 \theta_0)$. However, for the amplitude of the phase modulation, $\theta_0$, considered in this article  $J_0(2 \theta_0)\simeq 1$ \cite{footnote0}. As a result, the reasoning for the transition between bands can be performed on the bare lattice (with no modulation). In practice, only the first orders play a role since $J_p(x) \sim x^p$ when $x$ is about zero.

For a homogeneous lattice, the Bloch functions $u_{n,k=0}(x)$ et $u_{n+1,k=0}(x)$ of two successive bands at $k=0$ have opposite parities.  The last term of Eq.~(\ref{eqpotmodphas}) is responsible for a non zero matrix element between those two successive Bloch functions:
\begin{equation}
\langle u_{n,k=0} | V_{0, \theta}(x,t) | u_{n+1,k=0}\rangle  \neq 0.
\end{equation}
The corresponding interband transition is therefore allowed as a 1-phonon process.
Interestingly, the matrix element between bands having the same parity (still at $k=0$) does not vanish:
$$
\langle u_{n,k=0} | V_{0, \theta}(x,t) | u_{n+2,k=0}\rangle \neq 0.
$$
The first contributing term scales as $\cos(2\omega t)$ revealing the underlying 2-phonon process.

In summary, the phase modulation enables 1-phonon transitions to bands with opposite parity ($n=1\to n=2$, $n=1\to n=4$, ...), while the transition to bands with the same parity is allowed through a 2-phonon process.  

For amplitude modulation ($\varepsilon(t)=\varepsilon_0 \sin(\omega t)$) the strength of the transition is given by the matrix element of the modulating term $\varepsilon_0 \cos^2( \pi x/d)$ between the Bloch functions. Only 1-phonon transitions between bands of identical parity are therefore allowed. We stress that those selection rules are only valid for $k=0$ and for a potential having a discrete translational symmetry. In practice, when a Bose-Einstein condensate is loaded into the lowest band it is not uniquely projected on $k=0$, but has also components with the same weight on $-k$ and $k$ with $k \neq 0$. Furthermore, the extra harmonic confinement superimposed to the lattice breaks the discrete translational symmetry.

\section{Numerical simulations}
\label{sec:numerics}

\begin{figure}[t]
\centering
\includegraphics[width=\linewidth]{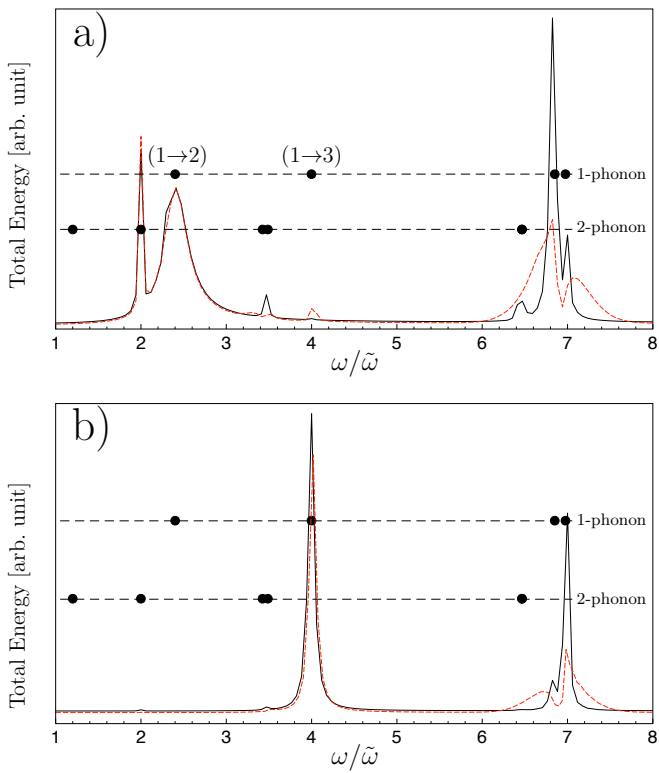}
\caption{Numerical results. The total energy associated with the Gross-Pitaevskii equation after the modulation is plotted as a function of the frequency of modulation. The numerics have been performed for an optical lattice of depth $s_0=4.86$ for phase  (a)  and amplitude (b) modulations ($\tilde \omega = 2\pi \times 6.55$ kHz). The solid line corresponds to simulations performed in the presence of repulsive interactions of strength $\beta=0.8$ (a value close to the experimental case, see text), and the red dashed line refers to similar simulations but in the absence of interactions ($\beta=0$). The 1 and 2-phonon transitions inferred from the band structure are represented by black disks.}
\label{fig0}
\end{figure} 

To confirm this qualitative analysis and conduct a first investigation of the role of interactions on the resonant excitations, we have performed numerical simulations of the 1D time-dependent Gross-Pitaevskii equation using the GPELab toolbox \cite{GPELab} which solves the dimensionless equation
\begin{eqnarray}
i\frac{\partial \psi}{\partial\tilde t}   & = &  \bigg[\frac{-\Delta + \tilde \omega^2_{E}X^2}{2} - \tilde s(t)\cos^2\left( \frac{\pi X}{4}+\tilde \theta(t)\right) \nonumber \\ 
&+ & \beta |\psi|^2 \bigg]\psi. 
\label{eqgpe}
\end{eqnarray}
The dimensionless time is normalized to $\tilde t = \tilde \omega t$ with $\tilde \omega^{-1} = md^2/(16 \hbar)=24.3$  $\mu$s for our parameters, $ \tilde \omega_{E}=\omega_{E}/\tilde \omega$, and $\tilde s_0=\pi^2s_0/8$. The value of the dimensionless interaction parameter $\beta$ is on the order of 0.8 for our experiments \cite{PRL2016}. In practice, we first determine the ground state through an imaginary time evolution for the static lattice potential of depth $s_0$. We then use this state as an initial condition to run the evolution of the wave function under the time-dependent lattice potential (Eq.~(\ref{eqgpe})). 

 In Fig.~\ref{fig0}, we plot the evolution of the total energy of the BEC for a fixed duration of the modulation time (100 in dimensionless unit) as a function of the normalized modulation frequency. The results are obtained for both phase (Fig.~\ref{fig0}a) and amplitude (Fig.~\ref{fig0}b) modulations, and in the absence ($\beta=0$, dashed line) or presence ($\beta=0.8$, solid line) of interactions. Black disks indicate the frequency of interband transitions at $k=0$ for 1 and 2-phonon transitions.

For phase modulation (Fig.~\ref{fig0}a), the main peaks coincide with the expectation from the selection rules. For the investigated range of modulation frequencies, the main peaks are: (1) the 1-phonon transitions from band 1 (ground state) to 2 ($\omega/\tilde \omega\simeq 2.41$), and from band 1 to 4 ($\omega/\tilde \omega\simeq 6.82$), and (2) the 2-phonon transition from 1 to 3 ($\omega/\tilde \omega\simeq 2$). Another less intense 2-phonon transition is observed at $\omega/\tilde \omega\simeq 3.53$ (from band 1 to 5). We also observe a few forbidden transitions: a small peak for the 1 to 3 transition ($\omega/\tilde \omega\simeq 4$) and a 1-phonon transition from 1 to 5 ($\omega/\tilde \omega\simeq 7.06$). We have shown numerically that the amplitude of those small peaks decreases when the external confinement is weakened. This confirms that we can ascribe the violation of the selection rules to the breakdown of the translational invariance by the external confinement. Comparing the $\beta=0$ and $\beta=0.8$ energy spectra, we find that the 2-phonon transition at $\omega/\tilde \omega\simeq 2$ is unaffected by interactions. We have also checked that this line is immune to the value of the external confinement. Those results are consistent with our previous studies dealing with the intrasite dipole mode known to be independent of the strength of interactions \cite{PRL2016,PRA2018}.  The phase modulation at this frequency simply excites this oscillatory mode. The interactions modify the other lines. For instance, the 1-phonon transition on the forbidden line at $\omega/\tilde \omega\simeq 4$ is washed out, or the 1-phonon transition on the 1 to 4 line ($\omega/\tilde \omega\simeq 6.8$) increases significantly and acquires a reduced width. Here the excitation drives the intrasite breathing mode. It means that such a line can be used a priori to monitor in situ the strength of the interactions. 

As expected from the parity argument, the energy spectrum for amplitude modulation exhibits lines with opposite selection rules for 1-phonon lines as compared to phase modulation. Similarly to the phase modulation results, we get a transition whose strength is strongly affected by interactions (1-phonon transition from band 1 to 5). The numerical simulations discussed in this section have been carried out with an amplitude of modulation similar to that used in experiments (see below). As intuitively expected, we have also seen numerically that the linewidth of the transitions increases with the amplitude of modulation. 
\section{Experimental spectra}
\label{sec:experiment}

\begin{figure}[t]
\centering
\includegraphics[width=\linewidth]{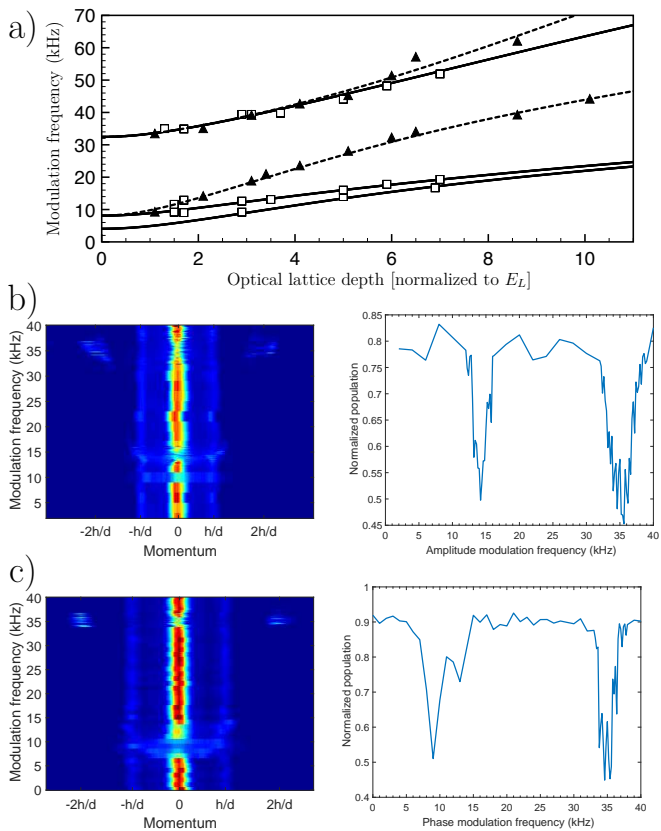}
\caption{(a) Summary of the resonances observed experimentally as a function of the lattice depth $s_0$ (normalized to the lattice characteristic energy $E_L$): for phase (open square) and amplitude  (filled triangles) modulations (modulation time equal to 3 ms). The solid (resp. dashed) line corresponds to the  allowed transitions with 1 or 2 phonons for phase (resp. amplitude) modulation.
(b) Sample set of experimental data for amplitude modulation for a lattice of normalized depth $s_0=2.1$. Left panel: each line of the image corresponds to an experiment performed for a given modulation frequency applied during 15 ms and subsequently imaged after a 25 ms time-of-flight. The experiment is repeated for a modulation frequency ranging from 2 to 40 kHz. Right panel: evolution of the zeroth order diffraction population as a function of the modulation frequency. (c) Similar experimental set of data for phase modulation at a lattice depth $s_0=1.7$. The 2 phonon line corresponds to the peak at the frequency $\sim 9.1$ kHz while the two other peaks are accounted for with a 1 phonon line.}
\label{fig1}
\end{figure}

As illustrated in Fig.~\ref{fig1}, the selection rules can be observed directly on the experiment. For this purpose, we proceed in the following manner. We load a BEC of $10^5$ rubidium-87 atoms into an optical lattice of depth $s_0E_L$ by ramping the laser intensity in 30 ms with a vanishing initial and final slope to avoid any excitation \cite{PRL2016,PRA2018}. The modulation is switched on 2 ms after this adiabatic loading and lasts for 1 to 5 ms for phase modulation and 3 to 15 ms for amplitude modulation. All confinement is subsequently turned off and an absorption image is taken after  a 25~ms time-of-flight. The maximum shift of the position of the lattice is $0.03d$ for phase modulation while the lattice depth varies as $s(t)=s_0(1+\varepsilon_0 \sin(\omega t))$ with $\varepsilon_0=0.07$ for amplitude modulation. The experiment is repeated for many modulation frequencies in the range of interest for a given lattice depth. The concatenation of those pictures are represented in Figs.~\ref{fig1}b and c for amplitude and phase modulation respectively. The image taken after the time-of-flight reveals the Fourier space components of the wave function. In the absence of excitation, we simply recover the well-known diffraction pattern of a BEC associated to a periodic structure of lattice spacing $d$ and that exhibits peaks separated by $h/d$  \cite{stringariinguscio}. For resonant excitations, atoms acquire a larger energy which results in a depletion of the zeroth order of the diffraction pattern and a concomitant increase of the populations of the other higher orders.

To process those images, we extract line by line the population in each diffraction order and infer from the depletion of the zeroth order the resonance frequencies. We find a good agreement between the experimental values and the ones calculated from the band structure spectrum as illustrated in Fig.~\ref{fig1}. We also observe directly the selection rules. Indeed, only 1-phonon lines are observed for amplitude modulation while one and two phonon lines can be excited by phase modulation. Note that, at low depth (1.5 and 1.8 $E_L$), we get an excitation for the phase modulation whose frequency coincides with an a priori forbidden line, a result ascribed to the role of the confining external potential as discussed in Sec.~\ref{sec:numerics}.

For rubidium atoms we do not benefit from an easily accessible Feshbach resonance without losses. We cannot therefore change the scattering length on a wide range and observe directly the effect of interactions on the excitation line as in the numerical results shown in Fig.~\ref{fig0}. However, the density can be tuned through the value $\omega_{E1,2}$ of the angular frequency associated to the external confinement (see Section \ref{sec:overview}). As a result, the lines can a priori be modified for two reasons: the compression/decompression of the trap and the interaction strength through the atomic density.

The projection of the ground state wave function on the Bloch functions of the lowest band depends a priori on the external confinement. By increasing the external confinement, the shape of the wave function is changed and departs more and more from the quasi-momentum $k=0$ eigenfunction \cite{InteractionBroadening}. As a result, the number of modes with quasi-momenta $k\neq 0$ of the ground band basis that should be used to get a complete representation of the wave function increases.  This spreading of the wave function on a larger $k$ extent suggests that the lines are sensitive to the interaction strength.

\begin{figure}[t]
\centering
\includegraphics[width=\linewidth]{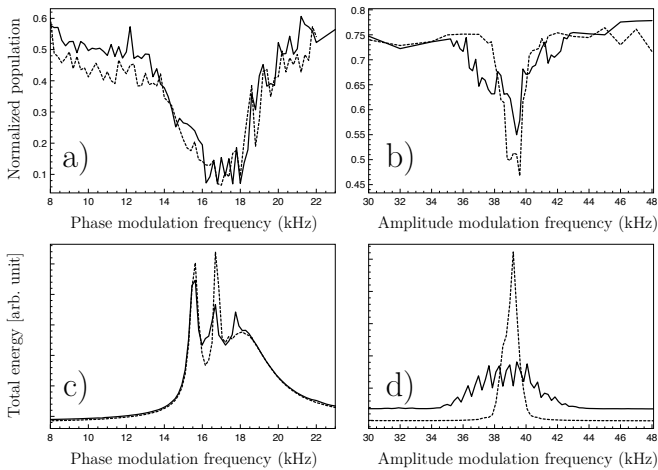}
\caption{(a) and (b): Experimental data: population in the zeroth order of the diffraction pattern obtained experimentally after a 25 ms time-of-flight as a function of the frequency of modulation: (a) 2-phonon line for phase modulation with a lattice depth $s_0=6.2$ (modulation amplitude $\theta_0=0.065$), (b) 1-phonon line for amplitude modulation with a lattice depth $s_0=3.1$ (modulation amplitude $\theta_0=0.035$). The solid line (resp. dashed line) corresponds to the compressed (resp. decompressed) trap. (c) and (d) variation of the total energy as a function of the driving frequency, obtained for numerical simulations performed in the same conditions as the experiment. }
\label{fig2}
\end{figure}

In figure \ref{fig2}, we have represented two different sets of spectrum data recording: about the 2-phonon line for phase modulation and about the second 1-phonon line for amplitude modulation. The experimental results are compared to numerical simulations performed under the same conditions where the total energy is plotted as a function of the modulation frequency for the same amplitude of modulation as the one used in experiments. The depth of the optical lattice is determined in each case by exciting the intrasite dipole mode as detailed in Ref.~\cite{PRA2018}, and therefore by a method independent from the modulation experiments.
For phase modulation (Figs.~\ref{fig2}a and c) the 1-phonon line (18 kHz), the 2-phonon line (15.8 kHz) and the 3-phonon line (16.7 kHz) are excited. They appear to be essentially immune to the variation of the external confinement. This result was expected since those lines are associated to the excitation of the intrasite dipole mode \cite{PRL2016,PRA2018}. In contrast, the second 1-phonon line for amplitude modulation is spectrally broadened when the external confinement is stronger. The reason lies in the fact that this excitation is coupled to higher modes such as the breathing mode that is sensitive to interactions.

\section{Dynamics of excited atoms}
\label{sec:dynamics}

\begin{figure}[t]
\centering
\includegraphics[width=\linewidth]{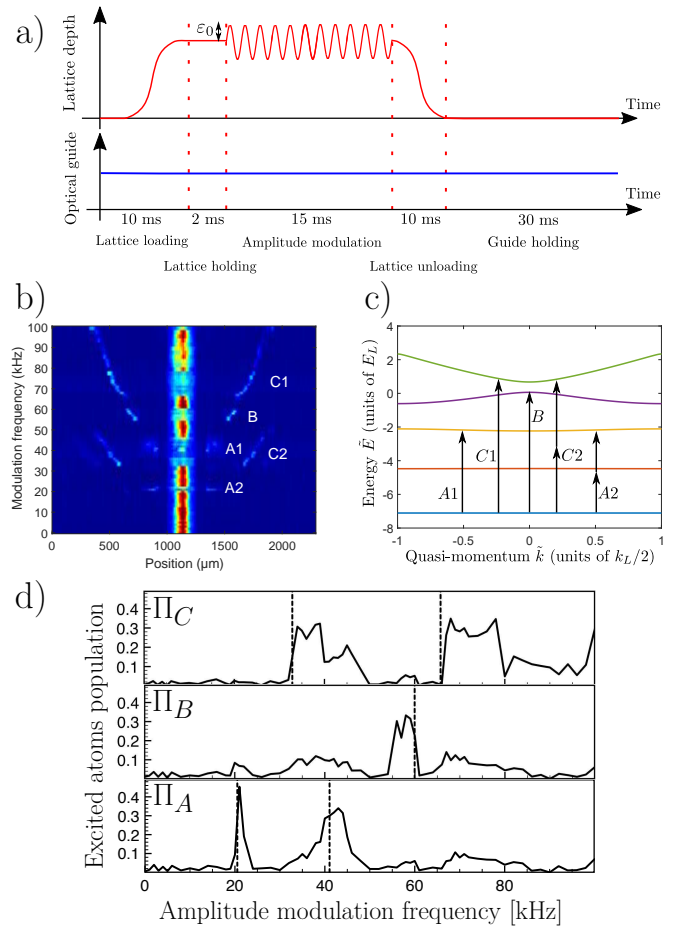}
\caption{a) Sketch of the experimental procedure to realize a mapping between the excited band through the modulation and the velocity space. b) Experimental images taken after applying the procedure a) and without time-of-flight. The amplitude modulation experiment has been here carried out for a lattice of depth $s_0=9$. c) Band structure associated with this lattice depth. The letters refer to the transition between the ground state and the excited bands. The number indicates the number of phonons absorbed in the corresponding excitation process. d) Populations of the excited atoms corresponding to a transfer to bands 3 (A), 4 (B) and 5 (C) as a function of the modulation frequency. The dashed lines indicate the 1 and 2-phonon transition frequencies at $k=0$: 20.53 kHz and 41.05 kHz for $A$, 59.95 kHz for $B$ and, 32.87 kHz and 65.75 kHz for $C$.}
\label{fig3}
\end{figure} 

We now focus on the dynamics of atoms once they have been excited to upper bands. Indeed, after an excitation for a non zero quasi-momentum the semiclassical local velocity is finite and given by the slope of the energy band:
\begin{equation}
v=v_L \frac{\partial \tilde E}{\partial \tilde k},
\end{equation}
where $v_L=h/md$ is the characteristic velocity of the lattice, $\tilde E$ is the energy normalized to $E_L$ and $\tilde k$ the wave vector normalized to $k_L=mv_L/\hbar$. For instance, the transition from band 1 (ground state) to band 3 (point $A1$ of Fig.~\ref{fig3}c) yields a maximum velocity on the order of a few hundreds of $\mu$m.s$^{-1}$. Interestingly, those velocities remain below the sound velocity and therefore do not deposit energy into the condensate because of the Landau superfluidity criterium. The size of the condensate being on the order of 200 $\mu$m, one shall wait for a very long time, a fraction of a second, to see in position space the excited atoms going out from the condensate. To overcome this limitation, we use the band mapping technique which amounts to accelerate the excited atoms. It consists in decreasing adiabatically the lattice intensity after the modulation \cite{bm1,bm2,bm3}. As a result, atoms lying on the first band (ground state) have a velocity ranging from 0 to $v_L/2$, those on the second band a velocity ranging from $v_L/2$ to $v_L$, and on the $n^{th}$ band a velocity belonging to the interval $[(n-1)v_L/2;nv_L/2]$ \cite{PRA2013}. With this mapping between velocity and the band number, we can identify the different kinds of excitation including the excitation through the absorption of two phonons. As a result of this adiabatic transformation, the atoms promoted by the modulation to higher bands, are accelerated and can collide in a dissipative manner with the BEC.

We have carried out such an experiment with an optical lattice of depth $s_0=9$ under an amplitude modulation. In this experiment, the modulation time is 15 ms and the amplitude modulation is large $\varepsilon_0=0.23$. As a result, we observe not only the allowed band transition at $k=0$ but also the ones at $k \neq 0$ that occur through 2-phonon transitions. Once the excitation is performed, we decrease the lattice and let the atoms evolve in the presence of the external confinement (see Fig.~\ref{fig3}a). To directly access to the mapping between band number and position, the images are taken without any time-of-flight. Results for each modulation frequency are summarized  on Fig.~\ref{fig3}b. 

We have identified the  excited bands using letters $A$, $B$ and $C$ followed by a number 1 or 2 designing the number of phonons involved in the excitation process (see Fig.~\ref{fig3}c). For instance, $A1$ corresponds to the 1-phonon transition from band 1 to 3, while $A2$ refers to the 2-phonon transition between the same bands. We indeed observe that $A1$ and $A2$ packets belong to the same range of velocities. The same conclusion holds for $C1$ and $C2$ between bands 1 and 5. The distance over which the packets propagate follows the expected hierarchy: atoms from packet A (band 3) have traveled over a shorter distance than atoms from packet B (band 4), and those of $B$ over a shorter distance than those of packet $C$ (band 5). This confirms the efficiency of the mapping procedure. Figure \ref{fig3}d provides the populations $\Pi_A$, $\Pi_B$ and $\Pi_C$ in the excited bands as a function of the modulation frequency extracted from the image Fig.~\ref{fig3}b. In practice, the populations $\Pi_i$ pertain to different velocity classes. To determine them as a function of time, we proceed each image in the following manner: we extract the maxima of all visible atomic clouds, and  integrate the image on areas of width 60 pixels about them. This procedure allows us to take into account the displacement of the atoms due to their velocity. The dashed lines indicate the transition frequency associated to the 1 and 2-phonon lines from the ground state band to the excited bands at $k=0$. Interestingly, we recover that the 2-phonon lines (for $A$ and $C$) have a width twice smaller than the 1-phonon line. For $A$ and $C$, the atoms are excited essentially above the frequency indicated by the dashed line since the atoms that are excited have a quasi-momentum $k\neq 0$, and the corresponding excitation frequency is larger than that at $k=0$ as a result of the positive curvature of the band about $k=0$.
 In contrast, the population in $B$ is excited for a lower frequency than that at $k=0$ because of the negative curvature of the band in the center of the Brillouin zone. We get a good understanding of the spectra of Fig.~\ref{fig3}d based on the 1 and 2-phonon lines.  However, it is not always easy to extract with a high reliability the populations. Indeed, some extra structures appear in the populations. Some of them may be interpreted as fingerprints of 3-phonon lines.  The method used to extract the populations also gives from time to time some overlaps of the integration areas, when atoms of different populations have velocity too close from each other. This limitation is related to non adiabatic effects near the edges of Brillouin zones, a well known limitation of the mapping band technique \cite{Natu}. 

We note that it is also delicate to extract the exact velocity with a reasonable accuracy since we do not know exactly at which moment atoms are promoted to the excited band during the 15 ms modulation time. Furthermore, atoms do not evolve freely. They are guided by an horizontal optical dipole trap that generates a harmonic potential whose quarter of period is on the order of 25 ms, a time close to the holding time in the guide.

In conclusion, we have investigated both experimentally and numerically the excitation of atoms in an optical lattice whose phase or amplitude is modulated. We have observed the selection rules and discussed their validity domain. Our study highlights the role of interactions even when they are relatively weak. The band mapping technique has been used to identify in more details the fate of the excited atoms.  This work opens up a few interesting perspectives for a deeper understanding of the role of interactions for Floquet engineering.
 
 This work was supported by Programme Investissements d'Avenir under the program ANR-11-IDEX-0002-02, reference ANR-10-LABX-0037-NEXT, and the research funding grant ANR-17-CE30-0024-01. M.~A. acknowledges support from the DGA (Direction G\'en\'erale de l'Armement).


\begin{thebibliography}{99}
\bibitem{footnote0} The expansion of the zeroth order Bessel function about zero reads $J_0(x)=1-x^2/4+O(x^4)$. For $\theta_0 = 5 \pi / 180$, we get  $J_0(2\theta_0)\simeq 0.9924$.
\bibitem{footnote1} $E_L=4E_R$ where $E_R=h^2/(2 m \lambda^2)$ with $\lambda=2d$ is the recoil energy associated to the absorption of a single photon from the laser producing the lattice.
\end{thebibliography}
\end{document}